\documentclass[prb,aps,twocolumn,showpacs]{revtex4}
\usepackage{graphicx}

\def\bbf{\bf }

\begin{document}

\title{Coulomb interaction and electron-hole asymmetry in cyclotron resonance of bilayer graphene
 in high magnetic field}
\author{V.~E.~Bisti$^{1,2}$, N.~N.~ Kirova$^2$}
\email{bisti@issp.ac.ru}
 \affiliation{$^1$Institute of Solid State
Physics, Russian Academy of Sciences, 142432, Chernogolovka,
Russia, $^2$LPS, CNRS and Universit\'e Paris-Sud, 91405, Orsay Cedex, France}

\date{\today}
\begin{abstract}
Inter-Landau-level transitions in the bilayer graphene in high
perpendicular magnetic field at the filling-factor $\nu=0$ have
been studied. The next-nearest-neighbor transitions, energy
difference between dimer and non-dimer sites, and layer asymmetry
are included. The influence of Coulomb interaction  is taken into
account. The magnetoplasmon excitations in bilayer graphene at
small momenta are considered within the Hartree-Fock
approximation. The asymmetry in cyclotron resonance of clean
bilayer graphene is shown to depend on magnetic field. At lower
magnetic fields the energy splitting in the spectrum is due to
electron-hole one-particle asymmetry while at higher magnetic
fields it is due to Coulomb interaction. For the fully symmetric
case with half-filled zero-energy levels the energy splitting
proportional to the energy of Coulomb interaction is found both
for bilayer and monolayer graphene.
\end{abstract}
\maketitle

\section{Introduction}

Recent experimental progress has allowed the fabrication and study
of monolayer and bilayer graphene.  The electronic band structure
of these objects is gapless and has a chirality \cite{0}. The
monolayer has Dirac-type spectrum with linear dispersion and
chirality exhibiting Berry phase $\pi$. The bilayer graphene is
the unique object which combines the parabolic dispersion  law of
quasiparticles near the zero energy point with their chirality
exhibiting Berry phase $2\pi$. This picture is obtained with the
tight-binding Hamiltonian for electrons  taking into account only
nearest-neighbor transitions; the one-electron spectrum  is
symmetric around zero energy.  Taking into account
next-nearest-neighbor transitions results in the asymmetry of
electron spectrum around zero-energy point \cite{Obzor}. Some
experimental data  demonstrate electron-hole asymmetry
 in cyclotron resonance spectra of clean bilayer graphene
\cite{113,Kim-02-09,Fogler,Kuzmenko,Milan-new} .

One-particle  Landau levels in the bilayer graphene at high
magnetic fields have been considered in the works \cite{110, 18}
taking into account only nearest-neighbor transitions. In magnetic
field there is a two-fold degenerate zero-energy Landau level
incorporating two different orbital states with the same energy.
Taking into account spin and valley degeneracies, the zero-energy
Landau level  in a bilayer is eight-fold degenerate. For the
bilayer with small asymmetry there are  four weakly split two-fold
degenerate  levels near zero energy. The valley and orbital
degeneracies are lifted, but the electron-hole symmetry is
preserved.

The near-zero-levels are strongly influenced by Coulomb
electron-electron interaction. The electron-electron interaction
is an important problem in the experimental study of cyclotron
resonance in monolayer \cite{36, 37}, bilayer \cite{113,Kim-01-10}
and multilayer \cite{4, 35, 20,21,22,Milan-new}
 graphene, exhibiting some properties
of a monolayer and bilayer. Interaction-induced shift of the
cyclotron resonance as a function of the filling-factor in bilayer
graphene was studied in Ref. \cite{Kim-01-10},the symmetry
breaking in the  zero-energy Landau level in bilayer graphene is
demonstrated  in \cite{Kim-02-10}.

   The charge-density excitations at small momenta
were considered theoretically within the Hartree-Fock
approximation for monolayer graphene \cite{16,112} and for bilayer
\cite{Bisti-Kirova}.  In the works \cite{11,34,Goerbig}
electromagnetic response in graphene was calculated numerically in
the RPA approximation for wide range of excitation momenta. The
spin-flip excitations and spin-waves in graphene were studied in
Ref.\cite{Goerbig-10}. In the works\cite{32, 32a} intra-Landau
level transitions were considered. The many-body corrections
obtained within the renormalization method, including weak
electron-hole asymmetry, and the attempts to explain sharp
transition from square to linear dispersion regime were reported
in Refs. \cite{Shizuya-09,Shizuya-10,Shizuya-11}. In the works
\cite{16, 112,Bisti-Kirova} Coulomb interaction was shown to
conserve electron-hole symmetry for excitations.

In the present paper the inter-Landau-level transitions in the
bilayer graphene in high perpendicular magnetic field at the
filling-factor $\nu=0$ are studied. The novelty of this work is
that the electron-hole asymmetry and Coulomb interaction are
included into consideration simultaneously. Special attention is
given to the difference in the cyclotron transition energies for
two valleys under different conditions.
 First, the
one-particle Hamiltonian and Landau levels are considered taking
into account the electron-hole asymmetry  due to
next-nearest-neighbor transitions and energy difference between
the dimer and non-dimer sites and the small energy difference
between the layers due to external potential. Then, the influence
of Coulomb interaction is included. The charge-density excitations
(magnetoexcitons) at small momenta are considered within the
Hartree-Fock approximation in the case of clean (neutral) bilayer
graphene with filling-factor $\nu=0$. The energies of excitations
are shown to be different in the two valleys, and the origin of
this difference depends on magnetic field.
 At lower magnetic fields the energy
splitting  is due to electron-hole one-particle asymmetry while at
higher magnetic fields the energy splitting in the spectrum is due
to Coulomb interaction. Next,the results are discussed in
connection with experimental possibility to observe the influence
of Coulomb interaction, and the comparison with other theoretical
works is presented.

\section{Hamiltonian and Landay levels of bilayer graphene}

The bilayer is modelled as two coupled hexagonal lattices with
inequivalent sites (A1, B1) and (A2, B2) in the first and second
graphene layers, respectively, arranged according to Bernal
(A2-B1) stacking. In the tight-binding model the energy states of
electrons in (A1-B2) dimer in the vicinity of zero-energy point
are conveniently described by an effective two-component
Hamiltonian
 \cite{110, 18,Obzor}
that operates in the space of wave functions $\Psi=(\psi_{A1},
\psi_{B2})$ in the valley $K$ and of $\Psi=(\psi_{B2}, \psi_{A1})$
in the valley $K'$. The asymmetry between on-site energies in the
two layers $U$ arising from the influence of external gates or a
doping effect, the next-nearest-neighbor transitions and the
difference between on-site energies of dimer and non-dimer sites
$\widetilde{\Delta}$  are taken into account{\bbf :}
\begin{equation}
H=H_0+H_1+H_2
\end{equation}
 \begin{equation}
 H_0=-\frac{1}{2m}
\left(
\begin{array}{cc}0& (\pi^+)^2\\
\pi^2& 0
\end{array}
\right)
\end{equation}

\begin{equation} H_1=\frac{\xi U}{2}
\left( \begin{array}{cc}1&0\\
0&-1
\end{array}  \right) -
\frac{1}{2m}\frac{\xi U}{\gamma_1}
\left( \begin{array}{cc}\pi^+\pi&0\\
0&-\pi\pi^+
\end{array}\right)
\end{equation}

\begin{equation}
H_2=\frac{1}{2m}\left(\frac{2\gamma_4}{\gamma_0}-\frac{\widetilde{\Delta}}{\gamma_1}\right)
\left( \begin{array}{cc}\pi^+\pi&0\\
0&\pi\pi^+
\end{array}\right)+\frac{\widetilde{\Delta}}{2}
\left(\begin{array}{cc}1&0\\
0&1
\end{array}\right)
\end{equation}

where $\pi=\hbar k_x+i\hbar k_y$, $\pi^+=\hbar k_x-i\hbar k_y$ are
the complex momentum operators, $\mathbf k$ is the wave vector
measured from the center of the valley, $\xi$ is the valley index,
$\xi=1$ in the valley $K$, $\xi=-1$ in the valley $K'$,
 $\gamma_0$ is the
intra-layer A-B coupling parameter, $\gamma_1$ is the inter-layer
 A2-B1 coupling parameter,
$m=\gamma_1/2v^2$ is the effective mass for bilayer graphene,
$v=\frac{\sqrt{3}}{2\hbar}a\gamma_0$, $a$ is the lattice constant.

The parameter $\gamma_4$ describes the next-nearest-neighbor
transitions (A1-A2 and B1-B2 interlayer hopping),
$\gamma_1=0.1\gamma_{0}$,  $\gamma_4=0.05\gamma_{0}$ (see
\cite{Fogler, Kuzmenko}).

$H_0$ is the basic term yielding a parabolic spectrum with the
effective mass $m$, and  its quasiparticles are chiral with the
degree of chirality related to Berry phase $2\pi$.

$H_1$ describes the layer asymmetry, leading to the opening of a
gap $\sim U$ in the spectrum.

$H_2$ is due to the next-nearest-neighbor transitions and the
difference between on-site energies of dimer and non-dimer sites.
The first term is responsible for the electron-hole asymmetry
 in the  spectrum around
zero-energy point. The second term due to $\widetilde{\Delta}$
results in the shift of the single-energy spectrum as a whole.
This term is unimportant for our considerations an will be omitted
later.

The two-component Hamiltonian is applicable if the considered
electron energy $\varepsilon|$ is within the energy range of
$|\varepsilon|<\frac{1}{4}\gamma_1$. The weak asymmetry means that
$U/\gamma_1\ll 1$, $\Delta/\gamma_1\ll 1$, $\gamma_4/\gamma_0\ll
1$.

In the perpendicular magnetic field $B$ the  energy spectrum of
Landau levels $E_{n\xi}$ ($n=0,1, \pm N$, $N=0,1,2...$) and
corresponding two-component wave functions $\Psi_{nk}$ are found
from the Hamiltonian $H$ (1) using the Landau gauge ${\mathbf
A}=(0, Bx)$ and raising and lowering operators
$a^+=l_B\pi^+/\sqrt{2}$ and $a=l_B\pi/\sqrt{2}$, as in the work
\cite{18}. The magnetic length $l_B$ and the cyclotron frequency
$\omega_c$ are defined as usual: $l_B=\sqrt{\hbar/eB}$,
$\omega_c=eB/m$, $e$ is the electron charge.  The basis consisting
of the wave functions describing the states in the ordinary
two-dimensional electron gas $\phi_{Nk}=e^{iky}\phi_{Nk}(x)$ is
used, where $k$ is the parameter which labels degenerate states
within one Landau level in Landau gauge.

\begin{equation}
E_{0}(\xi)=\frac{1}{2}\xi U,~~~E_{1}(\xi)=\frac{1}{2}\xi U
-\xi\delta
+(\frac{2\gamma_4}{\gamma_0}-\frac{\widetilde{\Delta}}{2\gamma_1})\hbar\omega_c
\end{equation}
\begin{equation}
 E_{\pm
N}(\xi)=\pm\hbar\omega_c\sqrt{N(N-1)}-\frac{1}{2}\xi\delta
+(\frac{\gamma_4}{\gamma_0}-\frac{\widetilde{\Delta}}{2\gamma_1})\hbar\omega_c(2N-1)
\end{equation}
where $\delta=U \hbar\omega_c/\gamma_1$.
$$
\Psi_{0k}=(\phi_{0k},0),~~~\Psi_{1k}=(\phi_{1k},0),
$$
\begin{equation} \Psi_{nk\xi}=(a_{n\xi}\phi_{Nk},b_{n\xi}\phi_{N-2,k})
\end{equation}

The coefficients $a_{n\xi}$ and $b_{n\xi}$ are the eigenvector
components.
$$
a_{n\xi}=1/\sqrt{1+D_{n\xi}},~~
b_{n\xi}=D_{n\xi}/\sqrt{1+D_{n\xi}}
$$
\begin{equation}
D_{n\xi}=\frac{E_{n\xi}-\xi U/2+\xi N \delta
-(\frac{\gamma_4}{\gamma_0}-\frac{\widetilde{\Delta}}{2\gamma_1})\hbar\omega_c(2N-1)}{\hbar\omega_c\sqrt{N(N-1)}}
\end{equation}
 Without any asymmetry in the zero approximation $a_{\pm
N,\xi}=1/\sqrt{2}, b_{\pm N,\xi}=\pm 1/\sqrt{2}$.

 The asymmetry splits the zero-energy
Landau level degenerate in valleys and orbital momenta into four
levels. Energy levels for $N\geq2$ are weakly split in valleys.

 Note that the spectrum of high-energy  LLs is applicable for the fields and levels satisfying the condition
$\hbar\omega_c\sqrt{N(N-1)}<\gamma_1/4$. For $\gamma_1=0.39 eV$
this inequality yields $B<50 T$ for $N=2$. For higher fields or
higher levels the full four-band Hamiltonian has to be used to
determine the exact LL spectrum \cite{5,33}.

The Zeeman splitting is omitted, and all levels are doubly
degenerate in spin. Although in graphite the electron $g$-factor
is not small ($g=2$), a very light effective mass $m \approx
0.054$ in the bilayer determines a small ratio between the Zeeman
energy and LL splitting $\varepsilon_Z/\hbar\omega_c\sim0.05$
\cite{18}. Trigonal warping coming from
$\gamma_3=\gamma_{A1-B2}\ll\gamma_1$ is not included.

\section{Coulomb interaction and magnetoexcitations}

 The total Hamiltonian of the many-body system in the perpendicular magnetic field with
 the Coulomb interaction is
 \begin{equation}
 \hat{H}=\sum E_{n\xi}a_{\lambda\xi\sigma}^+a_{\lambda\xi\sigma} + H_{int}
 \end{equation}
 where
$a_{\lambda\xi\sigma}^+$ and $a_{\lambda\xi\sigma}$
 are the one-particle creation and annihilation
operators; $\lambda=(n,k)$, $n=0, 1,\pm N$ indicates the Landau
level; $k$ is the parameter which labels degenerate states within
one Landau level in Landau gauge; $\xi$ and $\sigma$ are valley
and spin indexes.

\begin{equation}
H_{int}=\frac{1}{2}\sum V_{\lambda3,\lambda4}^{\lambda1;\lambda2}
a_{\lambda4\xi\sigma}^+a_{\lambda3\xi'\sigma'}^+a_{\lambda2\xi'\sigma'}a_{\lambda1\xi\sigma}
\end{equation}
The Coulomb interaction conserves spin and valley indexes;
$k1+k2=k3+k4$.

The matrix elements for Coulomb interaction are found using the
two-component wave functions (7), in analogy with calculations for
monolayer graphene \cite{16}.
\begin{equation}
V_{\lambda3,\lambda4}^{\lambda1;\lambda2}=V(q)e^{iq_x(k_1-k_2-q_y)}\tilde{J}_{n_4,n_1}(q)\tilde{J}_{n_3,n_2}(-q)
\end{equation}
\begin{equation}
 \tilde{J}_{m,n}(q)=a^*_{n}
a_{m}J_{|m||n|}({\mathbf q})+b^*_{n} b_{m}J_{|m|-2,|n|-2}({\mathbf
q})
\end{equation}
\begin{equation}
 J_{m,n}({\mathbf
q})=(\frac{n!}{m!})^{1/2}e^{-\frac{q^2}{4}}(\frac{q_y+iq_x}{\sqrt{2}})^{m-n}L^{m-n}_n(\frac{q^2}{2}),
\end{equation}
where $V(q)=\frac{2\pi}{\varepsilon q}$, $\mathbf
q=(q_x,q_y)$,$k4=k1+q_y$, $k3=k2-q_y$. $J_{m,n}({\mathbf
q})=J^*_{n,m}(-{\mathbf q}) (m>n$); $L^{m-n}_n$ are Laggerre
polinomials.

In Eq. (10)the summation is over the ensemble $n1,n2,n3,n4$, $k1,
k2$, both spins, both valleys and the wave vector $\mathbf
q=(q_x,q_y)$.

In this work only the charge-density-excitations (transitions
without changing the electron spin)  are studied, valley and spin
indexes $(\xi,\sigma)$ are not changed.  Corresponding operators
for excitations ($n,n'$) from the level $n$ to the level $n'$ with
the momentum $K$ are
 \begin{equation}
Q^+_ {n,n';\xi\sigma }(K)=\sum_k
a_{\lambda'\xi\sigma}^+a_{\lambda\xi\sigma}
 \end{equation}
 where $\lambda=(n,k)$, $\lambda'=(n',k+K)$.
It is assumed that the magnetic field is high which means that
$E_c\ll\hbar \omega_c$, where $E_c$ is the typical Coulomb energy:
$E_c=e^2/\varepsilon l_B$. The  momentum of excitation is small:
$Kl_B \ll 1$. The problem is considered in the way analogous to
that employed in \cite{KH} for the two-dimensional electron gas
with quadratic dispersion and in  \cite{16,112,Goerbig-10} for
monolayer graphene systems. The time-dependent Hartree-Fock
approximation is used. The Hartree-Fock approach assumes that
there is a small parameter $E_c/\Delta E_{nn'}(\xi)\ll 1$, where
$\Delta E_{nn'}(\xi)$ is the transition energy without
interaction.
\begin{equation}
\Delta E_{nn'}(\xi)=E_{n'}(\xi)-E_{n}(\xi)
\end{equation}
For monolayer graphene the ratio $\Delta E_{10}/E_c=2.77$
(see\cite{16}) and it does not depend on the magnetic field
strength due to linear dispersion. As stated in \cite{Goerbig-10},
the method works better for spin-flip excitations. For bilayer
graphene $E_c=10\sqrt{B}$, $\hbar\omega_{c}=2.2B$ (see \cite{110})
and the ratio $\hbar\omega_{c}/E_c=0.22B^{1/2}$ for $\epsilon=5$.
For the first high-energy transition
$E_{12}\simeq\sqrt{2}\hbar\omega_c$, and therefore for $B=40T$ the
ratio $E_{12}/E_c\simeq 2$. We do not consider with this method
the low-energy transitions between Landau levels 0 and 1 with
energies  close to zero.

The excitation energy $\widetilde{E}_{n,n';\xi\sigma}$ consists of
noninteracting and Coulomb parts:
\begin{equation}
\widetilde{E}_{n,n';\xi\sigma}=\Delta
E_{nn'}(\xi)+E^c_{nn';\xi\sigma}
\end{equation}

 Coulomb part $E^c_{nn',}$ is represented by the three terms: ``excitonic''
$E^{ex}_{nn'}$ part due to direct interaction of the electron at
the level $n'$ and the hole at the level $n$,  exchange
self-energy $\Sigma_{n\xi\sigma}$ and $\Sigma_{n'\xi\sigma}$
corrections to the one-electron Landau level energies and
``depolarization'' shift which is given  in the random phase
approximation (RPA). The RPA part is proportional to $K$, and is
important for dispersion (Ref.\cite{112}). However, we are
interested in the terms responsible for intervalley splitting
which are constants independent of $K$. Therefore the RPA part can
be omitted. This restriction enables to consider excitations with
different $(\xi,\sigma)$ independently.
 \begin{equation}
E^c_{n,n';\xi\sigma}=E_{n,n'}^{ex}+\Sigma_{n'\xi\sigma}-\Sigma_{n\xi\sigma}
\end{equation}

As for monolayer graphene, there is the problem of divergency of
exchange self-energy $\Sigma_{n}$ due to summation over all filled
LLs. The spectrum of monolayer and bilayer graphene described by
the model Hamiltonian in unbounded both from above and below. This
fact is physically artificial. In \cite{112} the cut-off value on
energy or number of LL vas defined. In \cite{16} the
semi-empirical way was used to treat the problem. The
electron-electron interaction parameter is fitted for one type of
transition to experimental data. For the bilayer graphene the area
of parabolic dispersion is less than required cut-off value, and
it is necessary to  consider the four-band Hamiltonian
\cite{Shizuya-09,Shizuya-10,Shizuya-11}.  In this work some
general rules are presented allowing to conclude where the Coulomb
part can be seen.

The interlayer electron transitions from the top filled (fully or
partially) to the next free (fully or partially) Landau levels
with energies nearly $\omega_c$  are considered. The selection
rules for these transitions are $\Delta N=1$.
 The case of filling-factor $\nu=0$ is considered. The Fermi level is equal to
zero. This filling-factor means the absence of charge density (the
absence of free carriers or the equal amount of holes and
electrons).  Different possible ground states in magnetic field
and different cyclotron transitions may correspond to this
filling.

\vskip 2mm
 {\bf A. The asymmetric bilayer without e-h asymmetry}.

Let $U>0$. In this case we have filling-factor $\nu=4$ for the
electrons in one valley and $\nu=4$ for the holes in another
valley. For the valley  with $\xi=1$ there is the top filled LL
with $n=-2$ and the transition (-2,1), and for the valley with
$\xi =-1$ there are the top filled 0 and 1 LLs and the transition
(1,2). Including spin there are two transitions of each type. The
noninteracting part is the same for both types of transitions:
\begin{equation}
\Delta E_{-2,1}(1)=\Delta
E_{1,2}(-1)=\omega_c\sqrt{2}+\frac{1}{2}|U-\delta|
\end{equation}

  Note that  the
electron-hole symmetry of one-particle Hamiltonian  leads to the
fact that (-2,1) and (1,2) transitions are really the same and
have the same energy. (-2,1) in electron representation is (1,2)
in hole representation. Taking into account spin degeneracy we
have four transitions with equal energies. Since all types of
asymmetry and electron-electron interaction are considered as
small perturbations, the many-body corrections can be calculated
with symmetric wave functions.


For $\nu=0$ integer filling, labelled $0I$,

$$
  E^{(1,2)}_{ex,0I}=-\frac{1}{(2\pi)^2}\int d{\mathbf
q}V(q)\tilde {J}_{22}({\mathbf q})\tilde{J}_{1,1}(-{\mathbf q})=
$$
\begin{equation}= \frac{1}{2(2\pi)^2}\int d{\mathbf
q}V(q)(J_{22}(q)+J_{00}( q))J_{11}(q))
\end{equation}

 $$(\Sigma_{2}-\Sigma_{1})_{0I} =\int
d{\mathbf q}\frac{V(q)}{2(2\pi)^2}(|J_{11}({\mathbf
q})|^2+|J_{10}({\mathbf q})|^2-
$$
$$-\int d{\mathbf
q}\frac{V(q)}{2(2\pi)^2}(\frac{1}{2}|J_{21}({\mathbf
q})|^2+\frac{1}{2}|J_{20}({\mathbf q})|^2)+
$$
\begin{equation}
 +\int d{\mathbf
q}\frac{V(q)}{2(2\pi)^2}\sum_{N=2}|J_{2,N}({\mathbf
q})J^*_{0,N-2}{\mathbf q}|
 \end{equation}
This value may depend on resolving the divergency problem, but it
is not zero. There is no Kohn's theorem \cite{14}.

 For the excitations in the different valleys  energy splitting
due to the layer asymmetry is absent.

 The same transitions may occur for the spin ferromagnetic state, where Zeeman
splitting would be included: one spin component $\sigma_1$ of LLs
0 and 1 is completely filled and the other $\sigma_2$ is
completely empty in both valleys. There are two excitations
$Q^+_{1,2;\xi\sigma_1}$ and two excitations
$Q^+_{-2,1:\xi,\sigma_2}$ with the same energies.

\begin{figure}[h]
\begin{center}
\includegraphics[width=8.5cm]{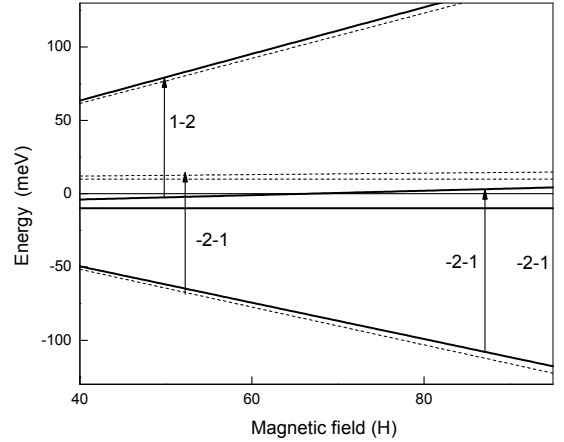}
\end{center}
\caption{ The one-electron energy levels and cyclotron transitions
taking into account  the next- nearest-neighbor transitions and
layer asymmetry (solid for K valley and dash for K' valley). }
\end{figure}
\vskip 2mm

 {\bf B. The asymmetric bilayer with electron-hole asymmetry}.

In this case at the total filling-factor $\nu=0$ the filling of
Landau  levels depends  of the magnetic field.

 If the magnetic field is not so high and
\begin{equation}
(\frac{\gamma_4}{\gamma_0}-\frac{\widetilde{\Delta}}{2\gamma_1})\hbar\omega_c-\delta<\frac{U}{2}
\end{equation}
then in the valley with $\xi=-1$ levels 0 and 1 are filled and in
the other valley with $\xi=1$ they are empty. There are cyclotron
electron or hole transitions (1,2) from the top filled to the next
empty level as for the case A (see Fig.1, left). The
noninteracting parts of transition energies are presented by the
following expressions:
\begin{equation}
\Delta E_{1,
2}(-1)=\hbar\omega_c(\sqrt{2}+\frac{\gamma_4}{\gamma_0}-\frac{\widetilde{\Delta}}{2\gamma_1})+\frac{1}{2}(U-\delta)
\end{equation}
\begin{equation}
\Delta
E_{-2,1}(1)=\hbar\omega_c(\sqrt{2}-\frac{\gamma_4}{\gamma_0}+\frac{\widetilde{\Delta}}{2\gamma_1})+\frac{1}{2}(U-\delta)
\end{equation}

The Coulomb parts are the same, as it was discussed before. The
difference in the energies between the valley transitions is due
to the electron-hole asymmetry.

\begin{equation}
\Delta E_{1,2}(-1)-\Delta
E_{-2,1}(1)=(\frac{2\gamma_4}{\gamma_0}-\frac{\widetilde{\Delta}}{\gamma_1})\hbar\omega_c
\end{equation}

If the magnetic field is sufficiently high and
\begin{equation}
(\frac{\gamma_4}{\gamma_0}-\frac{\widetilde{\Delta}}{2\gamma_1})\hbar\omega_c-\delta>\frac{U}{2}
\end{equation}
 then in both  valleys levels 1 are empty, but the level 0 is filled for the $K$ valley and empty for the $K'$ valley (see
 Fig.1,  right).
 There are only hole-type transitions (-2,1).
The following noninteracting parts of transition energies are
\begin{equation}
\Delta
E_{-2,1}(1)=\hbar\omega_c(\sqrt{2}-\frac{\gamma_4}{\gamma_0}+\frac{\widetilde{\Delta}}{2\gamma_1})+\frac{1}{2}(U-\delta)
\end{equation}
\begin{equation}
\Delta
E_{-2,1}(-1)=\hbar\omega_c(\sqrt{2}-\frac{\gamma_4}{\gamma_0}+\frac{\widetilde{\Delta}}{2\gamma_1})-\frac{1}{2}(U-\delta)
\end{equation}

Because of different filling of Landau levels in the valleys the
influence of Coulomb interaction differs in the part of the
self-energies.

\begin{equation}
\frac{1}{2}\Sigma_{20}-\Sigma_{10} =\int d{\mathbf
q}\frac{V(q)}{(2\pi)^2}(|J_{10}({\mathbf
q})|^2-\frac{1}{2}|J_{20}({\mathbf q})|^2)
\end{equation}
\begin{equation}
\Sigma_{10}-\frac{1}{2}\Sigma_{20}=\frac{7}{16}E_c\sqrt{\frac{\pi}{2}}
\end{equation}

\begin{equation}
\widetilde{E}_{-2,1}(1)-\widetilde{E}_{-2,1}(-1)=(U-\delta)+\frac{7}{16}E_c\sqrt{\frac{\pi}{2}}
\end{equation}

The splitting between valley cyclotron transitions is due to
Coulomb interaction and layer asymmetry

{\bf C. The bilayer graphene in the full-symmetric case}.

In this case labelled $h$ we have two half-filled zero-energy
levels in both valleys: 0 ($\nu_0=1/2$) and 1($\nu_1=1/2$); this
means $\nu=2$ for the electrons in each valley and $\nu=2$ for the
holes in each valley. For each valley  there are two transitions:
$(1,2)$ from half-filled to empty levels and $(-2,1)$ from filled
to half-empty levels. These transitions  are connected by the
Coulomb interaction $\tilde{V}$. Using the Hartree-Fock
approximation for non-integer filling-factors \cite{1,2} two
combined modes $Q^+_{s,a}$ with the energies $E_{s,a}$ are found.
\begin{figure}[h]
\begin{center}
\includegraphics[width=8.5cm]{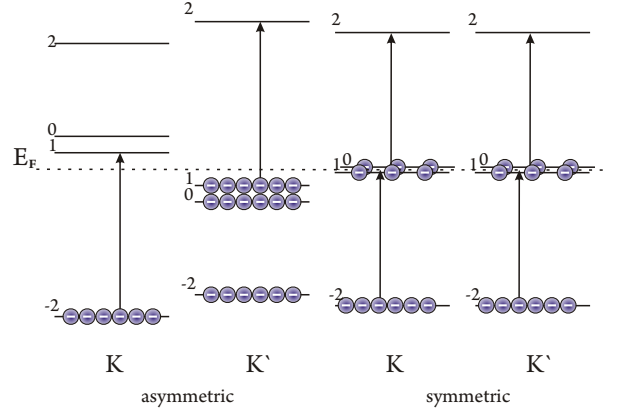}
\end{center}
\caption{The one-electron Landau energy levels and cyclotron
transitions for symmetric and asymmetric (B) cases. For the
symmetric case  there are two transitions in each valley.}
\end{figure}
$$
   Q^+_{s,a}=\frac{1}{\sqrt{2}}(Q^+_{1,2,\xi,\sigma} \pm Q^+_{-2,1,\xi,\sigma})
$$
\begin{equation}
E_{s,a}=\omega_c\sqrt{2}+E_{c}^{h} \pm \frac{1}{2}\tilde{V}
\end{equation}
These modes may be called  symmetric and antisymmetric in analogy
to modes in semiconductor bilayer.
\begin{equation}
E_c^{h}=\frac{1}{2}E^{(1,2)}_{ex,0I} + (\Sigma_2-\Sigma_1)_{h};~~
(\Sigma_2-\Sigma_1)_{h}\neq
\frac{1}{2}(\Sigma_{2}-\Sigma_{1})_{0I}
\end{equation}

$$
\tilde{V}=V_{-2,1}^{1,2}=V_{1,2}^{-2,1}=\frac{1}{(2\pi)^2}\int{d{\mathbf
q}V(q)\tilde {J}_{12}({\mathbf q})\tilde{J}_{-2,1}(-{\mathbf q})}=
$$
\begin{equation}
=\frac{1}{2(2\pi)^2}\int{d{\mathbf q}V(q)|J_{12}|^2({\mathbf
q})}=\frac{1}{2}\sqrt{\frac{\pi}{2}}E_c*\frac{7}{16}
\end{equation}

$\tilde{V}=2.5\sqrt{B}$ and for $B=40T$ $\tilde{V}\simeq 15 meV$.

This splitting for combined electron-hole transitions from
half-filled level is not specific to bilayer graphene. In
monolayer graphene with valley asymmetry considered in \cite{16}
the filling-factors were $\nu=2$ for electrons in one valley and
$\nu=2$ for holes in another valley. In the simple case with
symmetric valleys we have the half-filled zero-energy level in
both valleys which means $\nu=1$ for both electrons and holes in
each valley. For monolayer graphene with half-filled zero-energy
level there are related transitions $(0,1)$ for the electrons and
$(-1,0)$ for the holes, and the corresponding value of splitting
for the combined modes $\tilde{V}_{mg}=
\frac{1}{4}\sqrt{\frac{\pi}{2}}E_c$ is found using the wave
functions from \cite{16}. This value is nearly the same as for
bilayer graphene ($\tilde{V}_{mg}\simeq 2.5\sqrt{B}$), but for
monolayer graphene it is possible to observe this splitting for
lower experimentally used magnetic fields.

{\bf D. The symmetric bilayer with electron-hole asymmetry}.

 For the bilayer graphene with symmetric layers and therefore the
symmetric valleys, but with the electron-hole asymmetry included
there are half-filled 0 Landau levels ($\nu_0=1/2$)  and empty 1
Landau levels($\nu_1=0$) in both valleys. For each valley  there
are only hole-type $(-2,1)$ transitions. Coulomb corrections are
equal for both valleys due to the same filling, and there is no
splitting in this case.

\section{Summary and discussion}

In conclusion, the cyclotron transitions for clean bilayer
graphene are studied. The electron-hole asymmetry, the small layer
asymmetry and the influence of Coulomb interaction are taken into
account. The charge-density excitations at small momenta are
considered within the Hartree-Fock approximation.  It is shown
that the energies of cyclotron transitions in two valleys can be
equal or can be split either due to the electron-hole asymmetry or
due to Coulomb interaction. The splitting depends on applied
magnetic and electric fields.  At lower magnetic fields the energy
splitting in the spectrum  is due to electron-hole one-particle
asymmetry while higher magnetic fields the energy splitting in the
spectrum  is due to Coulomb interaction. For the fully symmetric
case with half-filled zero-energy levels (where electron-hole and
layer asymmetries are not taken into account) the energy splitting
proportional to Coulomb interaction is found both for bilayer and
monolayer graphene.

In the work  \cite{Shizuya-10}  the many-body corrections to
cyclotron resonance in monolayer and bilayer graphene due to
Coulomb interaction were studied neglecting the electron-hole
asymmetry. In the following work \cite{Shizuya-11} of the same
author the weak electron-hole asymmetry was included, where the
cyclotron resonances  were considered using the four-band
Hamiltonian. The four-band Hamiltonian consideration is more
precise, though the results concerning the noninteracting part
coincide in the limits of accuracy.
 The employed SMA method to study many-body corrections due to Coulomb interaction
(single mode approximation) for integer filling-factors is
identical to that used in the present work. Unfortunately it is
impossible to compare directly the results of these work with ours
in the aspect of many-body corrections because in
\cite{Shizuya-10,Shizuya-11} the different filling-factors
($\nu=4,8,12,16$) were considered.
 
In the experimental work \cite{Milan-new} graphene bilayers
embedded in the multilayer epitaxial graphene were studied.  The
splitting between electron-type and hole-type transitions is found
in a relatively narrow range of $B$. The possible explanation
invoking the electron-hole asymmetry yields the right values, but
should be seen for wide range of magnetic fields. Moreover, in
that work the other higher energy transitions corresponding to
$\Delta n=1$ were studied. In our notation the experimentally
studied transitions are (-2,3) and (-3,2). To compare the
experimental results with the theory it is necessary to study
these transitions in the future.

 In this work there are a simultaneous consideration of both
Coulomb interaction and electron-hole asymmetry, and suggestions
concerning observation of splitting due to these factors at
different regimes.

\section{Acknowledgements}

 Support is acknowledged to the Program LIA ENS-Landau, the Russian Foundation of Basic Research,
 project 10-02-00131, and to the ANR program in France (the project BLAN07-3-192276).

\end{document}